\documentclass[]{spie}  

\usepackage{amsmath,amsfonts,amssymb}
\usepackage{graphicx}
\usepackage[colorlinks=true, allcolors=blue]{hyperref}

\title{The MICADO first light imager for the ELT: first on-bench performance tests of the SCAO hybrid LQG+Integrator controller}

\author[a,b]{Nicolas LEVRAUD}
\author[a]{Florian FERREIRA}
\author[a]{Arnaud SEVIN}
\author[a]{Éric GENDRON}
\author[b]{Nicolas GALLAND}
\author[b]{Henri-François RAYNAUD}
\author[a]{Yann CLÉNET}
\author[c]{Richard DAVIES}
\author[b]{Caroline KULCS\'AR}
\affil[a]{LIRA, Observatoire de Paris, Université PSL, Sorbonne Université, Université Paris Cité, CY Cergy Paris Université, CNRS, 92190 Meudon, France}
\affil[b]{Université Paris-Saclay, Institut d’Optique Graduate School, CNRS, Laboratoire Charles Fabry, 91127, Palaiseau, France}
\affil[c]{Max Planck Institute for extraterrestrial Physics, Gießenbachstraße 1, 85748 Garching, Bayern, Deutschland}
\authorinfo{Nicolas Levraud: E-mail:nicolas.levraud@obspm.fr, Telephone: +337 88 33 01 02} 

\begin{document} 
\maketitle

\begin{abstract}
The Extremely Large Telescope (ELT) is expected to begin scientific operations in 2030, with MICADO as its first-light imager. The Single Conjugate Adaptive Optics (SCAO) system for MICADO is currently being assembled and tested at the Observatoire de Paris.
MICADO’s baseline AO control strategy uses a hybrid approach: a Linear Quadratic Gaussian (LQG) regulator for low-order modes, optimised to adapt to the disturbance caused by vibrations and ELT windshake, and an integral action controller for higher-order modes. Despite the non linearity of the Pyramid Wavefront Sensor (PyWFS), the so-called “optical gains”, LQG control has proven to be highly effective in simulation.

It has been shown that the LQG regulator is resilient to optical gain variations in the loop or to inaccuracies in their estimation. Increasing the number of modes controlled by the LQG regulator would therefore make the controller less sensitive to the optical gains, which are notoriously difficult to estimate during operation. A first hybrid controller, with only tip-tilt controlled with LQG control, has been validated using COMPASS, the end-to-end, GPU-accelerated, AO simulation platform.
We will present the first experimental validation of this hybrid controller on the MICADO-SCAO bench. We will also present results obtained by increasing the number of modes controlled with LQG control and the work carried out to go towards an autonomous LQG regulator.
\end{abstract}


\section{INTRODUCTION}
\label{sec:intro}  

With the increase in  of the ELT, the adaptive optics system are faced with new challenges. In particular the effect of the wind on the structures of the ELT creates a very strong Tilt signal detailed later. MICADO, as the first light imager of the ELT, will be among the first instruments to tackle this perturbation with its Single Conjugate Adaptiv Optics (SCAO). To prepare for this perturbation a more advance controller, the Linear Quadratic Gaussian (LQG) is included in the baseline for the Tip-Tilt mode. With the advancement of the integration of the MICADO SCAO bench, we can test on bench the performances of the baseline with the windshake in the perturbations. In this paper we present simulation results and the first on-bench results of the MICADO SCAO performances.  

\section{MICADO SCAO system and disturbances}

The MICADO AO is a SCAO, using M4 and M5 of the ELT for its wavefront correction correction, a Pyramid Wavefront Sensor (PyWFS) for measuring wavefront perturbations, and the COSMIC RTC (ref COSMIC) for control. The simulation is done using the COMPASS framework (ref compass) to make it compatible with the COSMIC system. The pyramid is measuring 
perturbations
at 700 nm (real bandwidth is between 600 and 950nm) with a 3$\lambda$/D modulation radius. We simulate the median conditions with a $r_0$ of 14.4cm (@500nm viewd at 30° zenith angle with the 35 layers ESO model) and a windspeed of 9.79m/s. The Point Spread Function (PSF) is computed at 2200nm with a time of integration of 10 seconds for Strehl ratio computation. The parameters are summarised in table \ref{tab:ELT_simu_condition}. 

\begin{figure}
    \centering
    \includegraphics[width=0.9\linewidth]{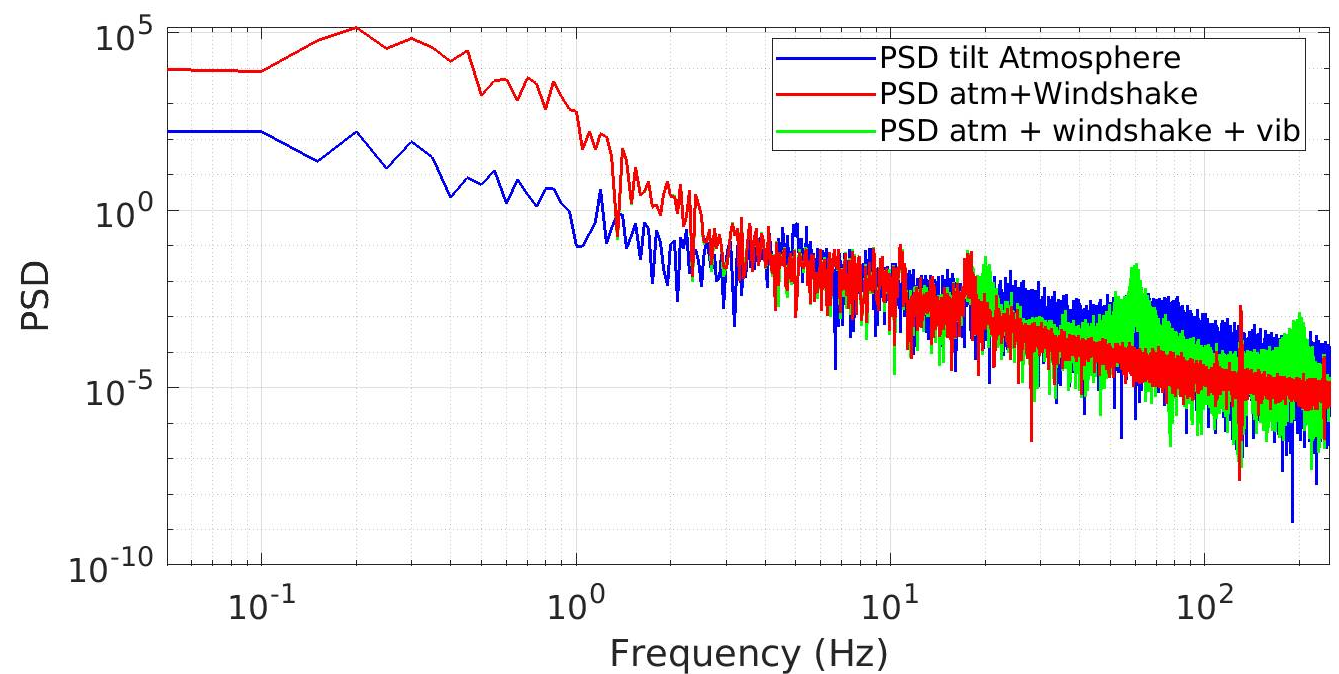}
    \caption{Comparison of PSD between atmospheric Tilt, atmosphere + windshake Tilt and atmosphere + windshake + vibration Tilt. }
    \label{fig:DSP_TT}
\end{figure}

Two more perturbations are added on top of the atmospheric turbulence: windshake as predicted by ESO, and vibrations. The windshake trajectory was simulated by ESO based on a  mechanical analysis of the effect of wind on the ELT structure. The resulting phase disturbance is a windshake Tilt with a 10x larger amplitude than the atmospheric Tilt. This effect can be seen on the Power Spectral Density (PSD) plotted in figure \ref{fig:DSP_TT}, compared with the atmospheric Tilt for the median conditions. Because of the presence of a wide low-frequency plateau (up to 5Hz in this case) in its PSD, windshake cannot be efficiently compensated by an integrator controller (see Conan et al 2011\cite{conan_are_2011}).

With the complexity and number of moving parts of the ELT, vibrations are to be expected -- like on most telescopes. In order to avoid amplifying high-frequency vibrations, the integrator gain should be set to a low value, which degrades overall performance.
To compare our control solutions vibration is added to atmosphere and windshake on the tip-tilt modes. From previous experience on different telescope we add vibrations with 3 main peaks at 20, 60 and 200Hz using a digital filter to produce this signal. The total amplitude is 45mas rms. The vibration's DSP on top of atmosphere and windshake is shown in Figure \ref{fig:DSP_TT}. These vibrations are in the amplification band of the integrator and force the integrator gain to be reduced, wich limits its performances. The LQG controller adapts the control rejection to these perturbations to compensate them efficiently, as shown in Sivo et al 2014\cite{sivo_first_2014}.

\begin{table}[]
    \centering
    \begin{tabular}{|c|c|}
        \hline
        $r_0$ & 14.4cm at 500nm (median condition) \\
        \hline
        Atmospheric turbulence model & 35 layers ESO model, 30°zenith angle \\
        \hline
        windspeed & 9.79 m/s \\
        \hline
        star magnitude & from 9.25 to 17.25 mag \\ 
        \hline
        Deformable Mirror & M4 + Tip-Tilt M5 \\
        \hline
        Wavefront Sensor & PyWFS at 700nm, $3\lambda/D$, OG measured with Esposito 2020 \\
        \hline
        PSF & measured at 2200nm \\
        \hline
        windshake & ESO windshake data package \\
        \hline
        vibration & 42 mas rms in total,vibration peaks $@$20,60 and 200Hz \\
        \hline
    \end{tabular}
    \caption{Condition of the ELT simulation}
    \label{tab:ELT_simu_condition}
\end{table}

\section{MICADO control baseline scheme} 

For MICADO, the baseline controller beeing implemented in COSMIC combines a leaky integrator for high order modes and a LQG controller for low order modes. As of this paper, low order modes only contain the Tip-Tilt, but it would be easy to compensate more modes with the LQG if needed.

\subsection{Leaky integrator} 

For each mode, the leaky integrator update equation at iteration $k$ is
\begin{equation}
    U_{k}=\alpha U_{k-1} + g\times y_k  \,,
    \label{eq:integrator}
\end{equation}
where $U_k$ is the modal command, $g$ is the modal gain, $y_k$ is the modal residual measured by the PyWFS and $0 < \alpha< 1$ is the leak factor.
The modal gain $g$ is optimized separately for each mode by the CLOSE algorithm (presented in \cite{deo_correlation-locking_2021}). CLOSE is not used in simulation but is necessary in bench operation. The integrator gain for Tip-Tilt (TT) is optimised by hand.

Due to the non linearity of the PyWFS, we need to perform optical gain compensation. We measure the optical gains of the PyWFS using the method presented in Esposito et al 2020\cite{esposito_-sky_2020}.
The integrator is particularly susceptible to optical gain measurement error as it is equivalent to a change in the integrator gain. Optical gain compensation are included in the measurement step so are already accounted for in $y_k$.

We will test the effect of a bad optical gain estimation in simulation and on bench by multiplying the optical gain for Tip-Tilt by a varying factor.

\subsection{Tip-tilt LQG} 

The LQG controller uses a data driven model of the atmospheric perturbation to predict the next iteration's perturbation. This model is obtained for each mode in the LQG controller using the method described in Sinquin et al 2020\cite{sinquin_-sky_2020}. This gives us the first parameter of the LQG: the order of the perturbation model. More complex perturbations requires a higher order of model to fit accurately the perturbation PSD. We either measure the modal Pseudo-Open Loop (POL) trajectories on a first closed loop using an integrator, or subsequently update our model using the loop telemetry. That way our model will adapt to any change of the atmosphere. We then construct a Kalman Filter from the identified disturbance model and the estimated measurement noise variance. In the construction of this kalman filter we find the second parameter of the LQG controller : the fudgefactor. This parameter modifies the value of the measurement noise variance used to compute the Kalman filter. This adjustment enables to make the system more robust to modeling errors (such as non-linearities). Finally, the modal command is computed as a two-steps ahead prediction of the disturbance. The resulting LQG controller is put in standard state-space form for implementation in the MICADO RTC. An example of LQG construction using the same formalism is shown in Marquis et al 2024\cite{marquis_first_2024}.

As a method of control using a linear prediction model, the LQG's behaviour in presence of optical gains is a major concern. One of the aim of this paper is to test the response of LQG to bad OG estimation. The OG compensation has been taken into account in the Kalman Filter computation following ref AO4ELT. 

\section{Simulation results}\label{sec:Simulation_result} 

\subsection{Controller Comparison}

\begin{figure}
    \centering
    \includegraphics[width=0.8\linewidth]{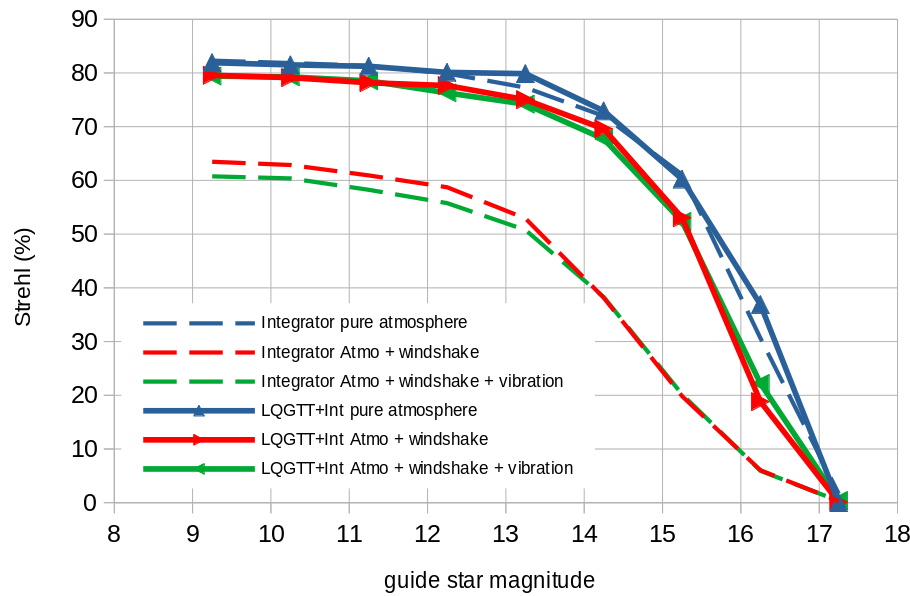}
    \caption{Performance of MICADO simulation with LQG VS integrator}
    \label{fig:magnitude_simulation_integ_LQG}
\end{figure}

Parameters of simulation are summed up in table \ref{tab:ELT_simu_condition}. We simulated the parameters for median atmospheric conditions for magnitude 9.25 to 17.25 first with pure atmospheric turbulence, then with windshake added, and finally with windshake and vibrations added. The results are shown on figure \ref{fig:magnitude_simulation_integ_LQG} with the dotted line. We see that windshake and vibration greatly degrade the residuals for every magnitude.

We simulated with the same parameters but using the MICADO baseline with LQG controller for tip-tilt and integrator for the other modes (LQGTT+Int). Thei ntegrator gains for all high-order modes were kept the same for the TT and LQGint+TT tests.
We see that in general the LQG restores performance close to the pure atmosphere case despite the added perturbations. The TT perturbation model orders were taken as follows : 5 for the pure atmosphere, 15 for atm + windshake, 25 for atm + windshake + vibrations.

\subsection{Optical gain effects} 

Optical gains are a problem specific to the PyWFS and a cause of concern  for any system using this sensor. As such we previously tested the effect of optical gains on the LQG controller, and in particular how a bad OG estimation affects the results. For Integrator a variation of optical gain is equivalent to changing the integrator gain, which heavily impacts performance. For the LQG it is less obvious. We probed that issue by multiplying the estimated OG on Tip-Tilt by a parameter between 0.5 and 2. 

\begin{figure}
    \centering
    \includegraphics[width=0.7\linewidth]{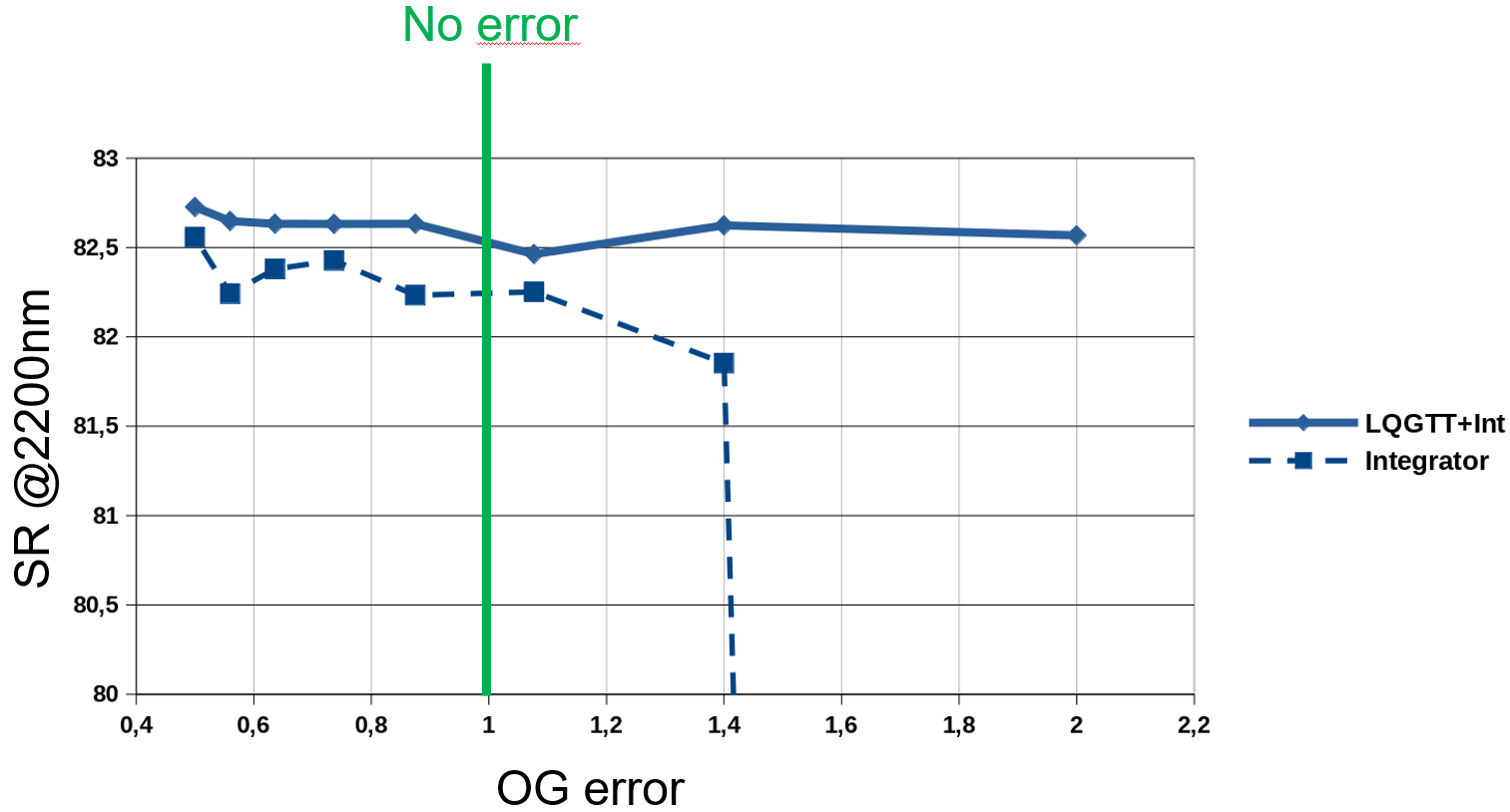}
    \caption{Effect of bad optical gain estimation in simulation No error green bar corresponds to the OG measured with no modification}
    \label{fig:OG_effect_simulation}
\end{figure}

The change in performance for both the integrator and LQG (in simulation) is plotted on Figure \ref{fig:OG_effect_simulation}.
We see that the method we use to estimate the OG probably overestimate OG on Tip-Tilt, since the integrator performs better when the OG is lowered. This appears to have very little impact on LQG performance.

\section{MICADO SCAO: first on-bench control performance tests} 
During its Assembly, Integration and Testing (AIT) phase, the MICADO SCAO will be available for testing in 2 different forms. It is currently in the `flat configuration' where it is assembled on optical tables. In the future it will be assembled in the MICADO SCAO carbon bench.

\subsection{MICADO flat configuration} 

 In the flat conf the bench gets its light through a telescope and atmospheric simulator using rotating phase plates to simulate atmospheric turbulence. The main difference is the deformable mirror used, a 64x64 ALPAO DM and a slow TT which simulate M4 and M5 of the ELT respectively. The bench parameters are summed in table~\ref{tab:bench_parameters}.

\begin{table}[]
    \centering
    \begin{tabular}{|c|c|}
        \hline
         $\lambda_{WFS}$ & large band 600 to 950nm  \\
        \hline
        $\lambda_{PSF}$ & 1550nm (laser) \\
        \hline 
        windspeed & 9.8m/s \\
         \hline 
        $r_0$ & 11.7cm (@500nm measured on bench) \\
         \hline 
         $\sigma_{NCPA}$ & 120nm (estimated through phase diversity)  \\
         \hline 
        DM & ALPAO 64x64 + slow TT \\
         \hline 
        Pupil & ELT (with old/large spiders) \\
        \hline
         Windshake & Amplitude adapted to measured Tilt (same proportion between bench\\ & tilt and added windshake as in simulation) \\
         \hline
         Vibration & peaks at 20, 60 and 200 Hz, amplitude adapted on measured Tilt \\
         \hline
    \end{tabular}
    \caption{on-Bench AO parameters}
    \label{tab:bench_parameters}
\end{table}

In the current configuration the windshake and vibrations need to be added by the phase corrector as additional tip-tilt perturbations. The windshake has a large amplitude but slow evolution, so it is added using the slow TT, while vibrations are added using the DM64x64. Furthermore, correction of the TT must be split between both actuators, since otherwise the correction would saturate the DM64x64, or the slowTT would not correct fast vibrations correctly. 

To tackle this issue a modification of the LQG and integrator controller has been implemented for these bench tests. On the real ELT, ESO will make the M4/M5 split after modal commands are computed.

\subsection{TT temporal separation} 

\begin{figure}
    \centering
    \includegraphics[width=0.6\linewidth]{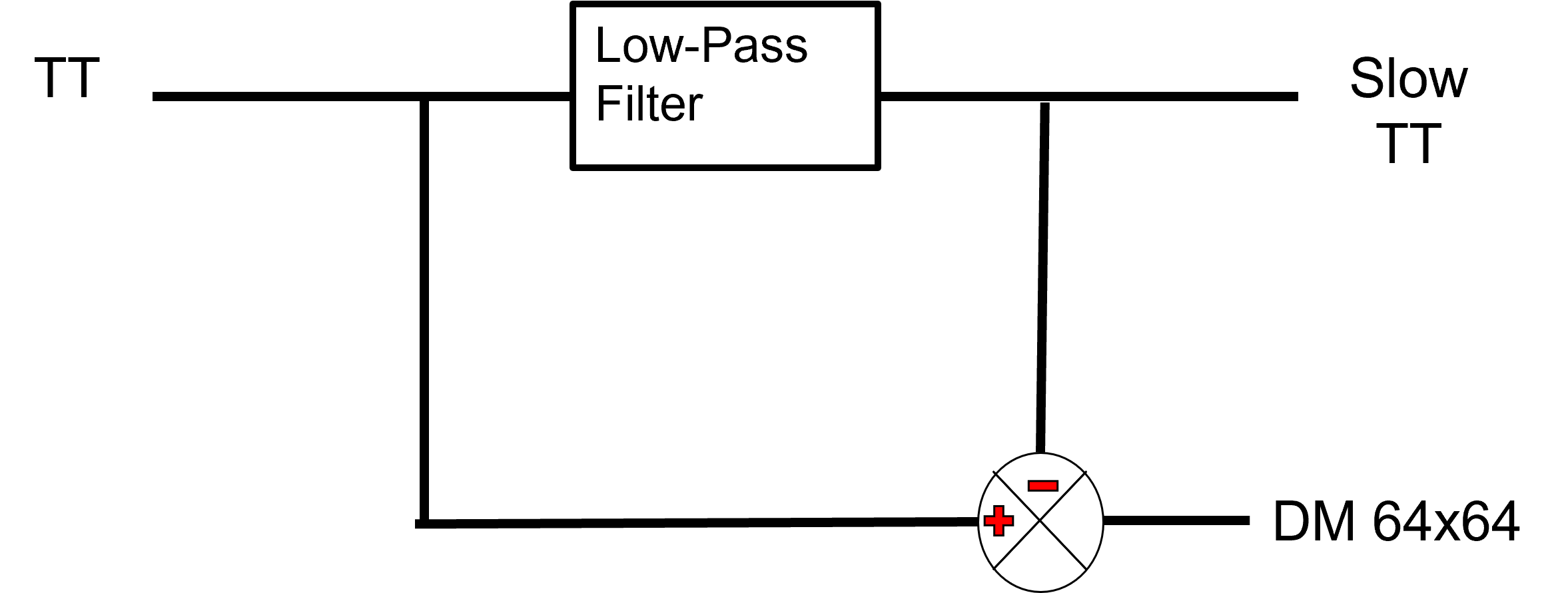}
    \caption{Separation between low frequency and high frequency Tip-tilt. Low frequency is sent to the slow TT and the remaining Tip-tilt is sent to the DM 64x64}
    \label{fig:2-DM_scheme}
\end{figure}

We want to modify the controllers we are using to add at the end a separation between the fast evolving TT and the slow evolving TT. In order to achieve this, we expand the state-space representation of the controller model by adding a low-pass filter which splits the modal command into a slow-varying and a fast-evolving components. These two tip-tilt commands are then which sent respectively to the slowTT and to the DM64x64 actuators, as shown by the block-diagram in figure \ref{fig:2-DM_scheme}. 
The low bandpass filter implemented is of the second order. It is defined by the recursive equation
\begin{equation}
    \alpha_0 z_k + \alpha_1 z_{k-1} + \alpha_2 z_{k-2} = u_k \,.
\end{equation}

To tune this low band pass filter the estimation of the band pass of the slowTT was necessary. This estimation was necessary since the slowTT manufacturer gave characteristics before the mirror was glued on top of it and consequently were not acurate anymore.

\begin{figure}
    \centering
    \includegraphics[width=0.30\linewidth]{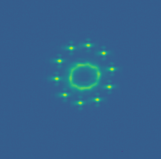}
    \includegraphics[width=0.55\linewidth]{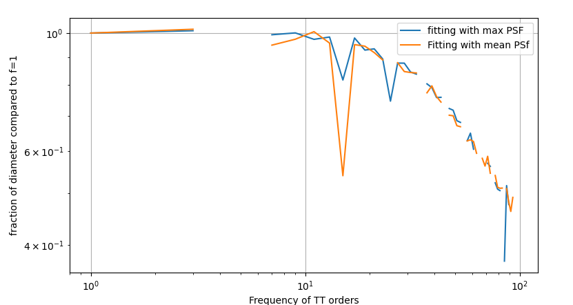}
    \caption{On the left we see the PSF for two frequency summed : outside circle is for 1Hz,inside circle is for 91Hz. Right is the plot of the results. It can be hard to fit the circle, so not every frequency gives results. }
    \label{fig:fcut_test}
\end{figure}

To perform this estimation we send a sine and cosine signal to tip and tilt with no atmosphere or perturbation resent. We measure the position of the PSF and the diameter of the circle it follows. In particular the evolution of said diameter with a changing frequency of the sine signal. The results are shown on Figure \ref{fig:fcut_test}

\begin{figure}
    \centering
    \includegraphics[width=0.45\linewidth]{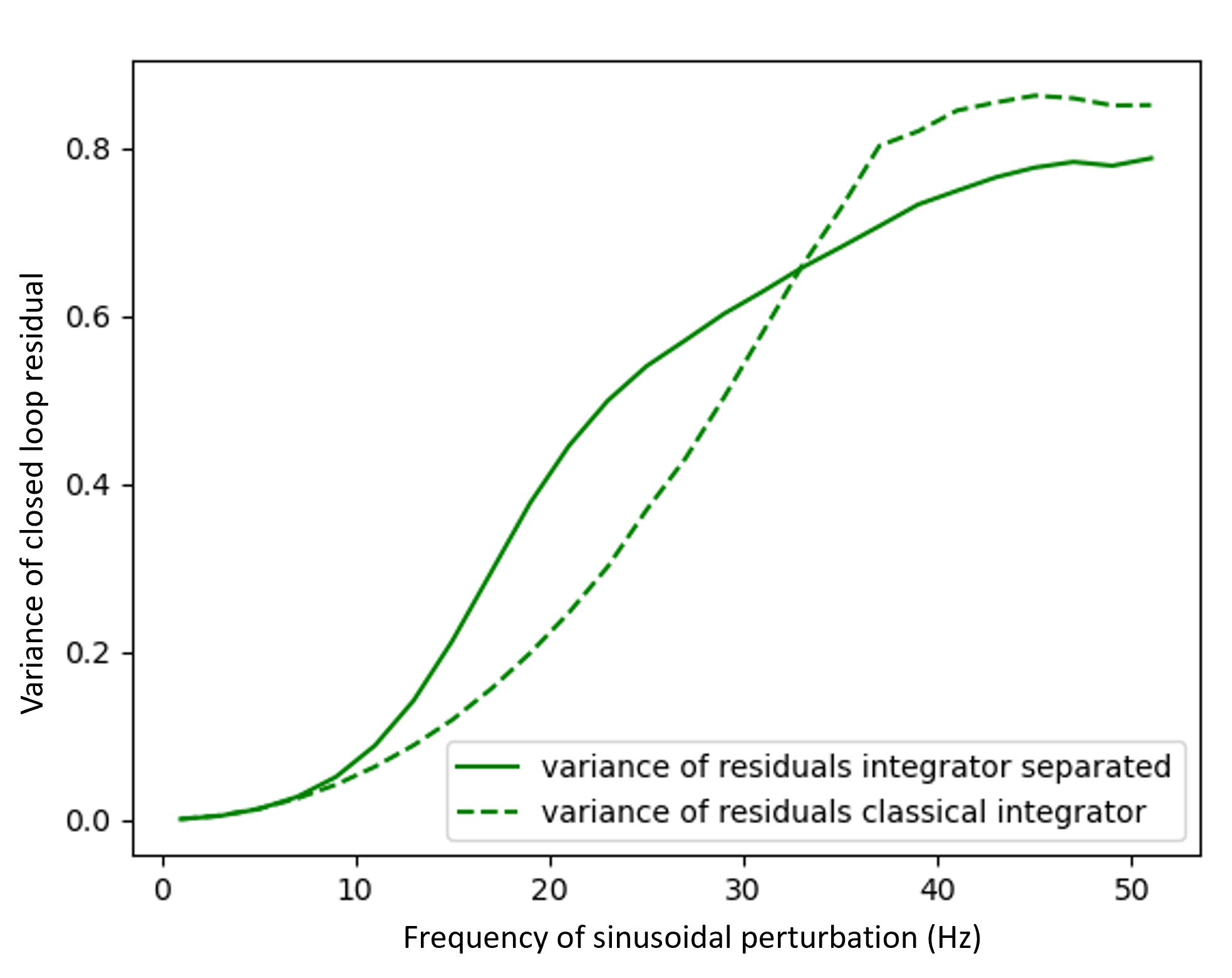}
    \includegraphics[width=0.45\linewidth]{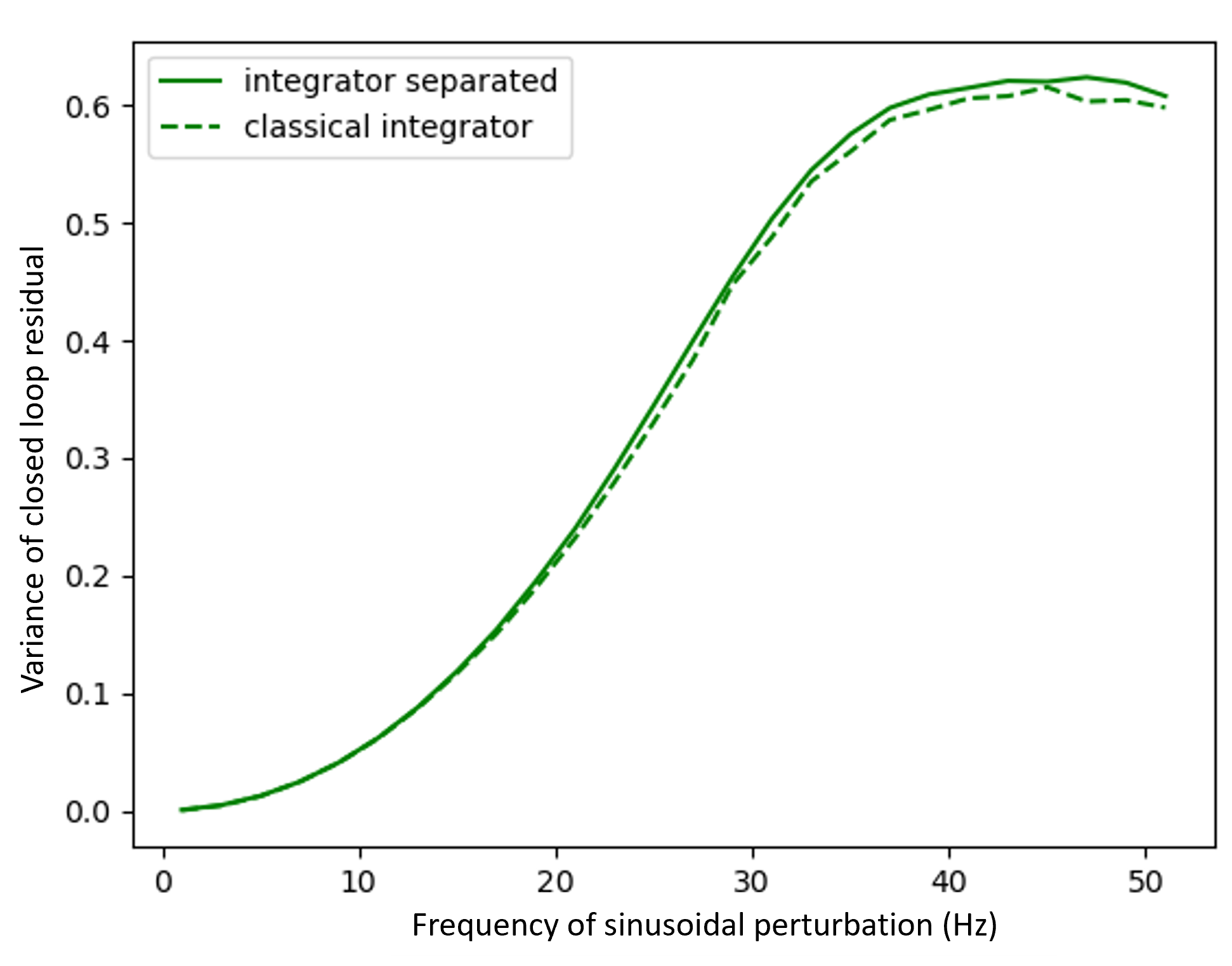}
    \caption{residual after closed loop with a sinusoidal tip tilt as perturbation. The aim is for the residuals with separation (full line) to give same results as with the classical integrator (dashed line). \textbf{Left:} using separation with 30Hz cutoff frequency.  \textbf{Right:} using separation with 10Hz cutoff frequency.}
    \label{fig:fcut_residualtest}
\end{figure}

Another test was to close the Loop on sinusoidal parameters and test with different cutoff frequency to measure the residuals and compare the pure integrator with the separated integrator on Figure \ref{fig:fcut_residualtest}. We check this way that the 30Hz cutoff frequency was optimistic as seen by the difference between the pure integrator (dashed line) and the separated integrator (full line). A 10Hz cutoff frequency is more adapted as we see that both line almost overlap (this test is done on real bench data, not simulated). 

A state-space realization of the block-diagram in figure \ref{fig:2-DM_scheme} is 
\begin{align}
x_{F,k} & = A_{mode}x_{F,k-1} + B_{mode}u_{TT,k} \,, \\
u_{S,k} & = C_{mode}x_{F,k} \,,
\end{align}
where $u_{TT}$ is the desired tip-tilt command, $u_S$ is the vector of the slowTT and DM64x64 commands, and
\begin{equation}
 A_{mode}=\begin{pmatrix}
    -\alpha_1 & -\alpha_2 & 0\\
    1 & 0 & 0 \\
    0 & 0 & 0 \\
    \end{pmatrix}
    \,\,\,\,\,
    B_{mode} = \begin{pmatrix}
    {1\over \alpha_0} \\ 0 \\ 1
    \end{pmatrix} \,\,\,\,\,
      C_{mode}=\begin{pmatrix}
    1 & 0 & 0\\
    -1 & 0 & 1
    \end{pmatrix} 
\end{equation}

Assuming that the modal controller is implemented in the standard state-space form
\begin{align}
x_{R,k} & = A_1 x_{R,k-1} + B_{1}y_{TT,k} \,, \\
u_{TT,k} & = C_1 x_{R,k} \,,
\end{align}
where $y_{TT}$ is the modal measurement, the state-space form of the modal controller plus M4/M5 split is
\begin{align}
x_{RS,k} & = A_3 x_{RS,k-1} + B_3 y_{TT,k} \,, \\
u_{S,k} & = C_3 x_{RS,k} \,,
\end{align}
where
\begin{align}
A_2=\begin{pmatrix}
    A_{mode} & 0 \\
    0 & A_{mode}
    \end{pmatrix} \,\,\,\,\,
  B_2=\begin{pmatrix}
    B_{mode} & 0 \\
    0 & B_{mode}
    \end{pmatrix}
    \,\,\,\,\,
    C_2=\begin{pmatrix}
    C_{mode} & 0 \\
    0 & C_{mode}
    \end{pmatrix} \,,
\end{align}
\begin{align}
A_3= \begin{pmatrix}
    A_1 & 0 \\
    B_2 C_1 A_1 & A_2
    \end{pmatrix} \,\,\,\,\,
  B_3=\begin{pmatrix}
    B_1 & -B_1\\
    B_2 C_1 B_1 & -B_2 C_1 B_1
\end{pmatrix}
    \,\,\,\,\,
    C_3= \begin{pmatrix}
    0 & C_2
 \end{pmatrix} \,.
\end{align}
    


\subsection{Bench performance metric}

The main metric used here is the Strehl ratio. It is estimated  with the formula in equation \ref{eq:SR} : 
\begin{equation}
    SR= {max(PSF_{\phi})\over max(PSF_{ref}) T_\phi}
    \label{eq:SR}
\end{equation}

\begin{figure}
    \centering
    \includegraphics[width=0.435\linewidth]{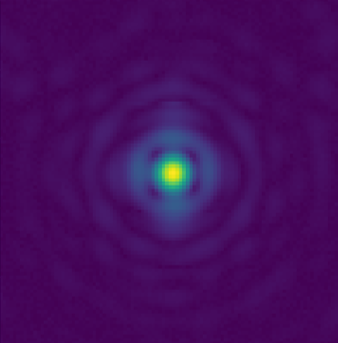}
    \includegraphics[width=0.45\linewidth]{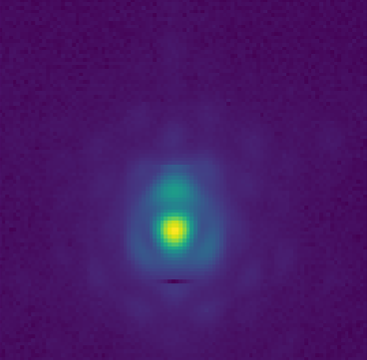}
    \caption{Exemple of on-bench PSF (sqrt scaling). \textbf{Left :} reference PSF with NCPA compensated. \textbf{Left :} best closed loop PSF with 31$\%$ Strehl  }
    \label{fig:PSF example}
\end{figure}

The transmission of the phase plates is estimated to be $T_\phi=0.85$, due to 2 phase plates being used. 
Example of the reference PSF is shown on figure \ref{fig:PSF example}. This reference PSF is measured with the phase plates out of the beam and with NCPA compensation. Example of a 31\% SR PSF is shown on the same figure. Due to the bench parameters the maximum SR reachable is 35\%. Residual petal modes are probably the source for the `trefoil' looking light ring.

\begin{figure}
    \centering
    \includegraphics[width=0.7\linewidth]{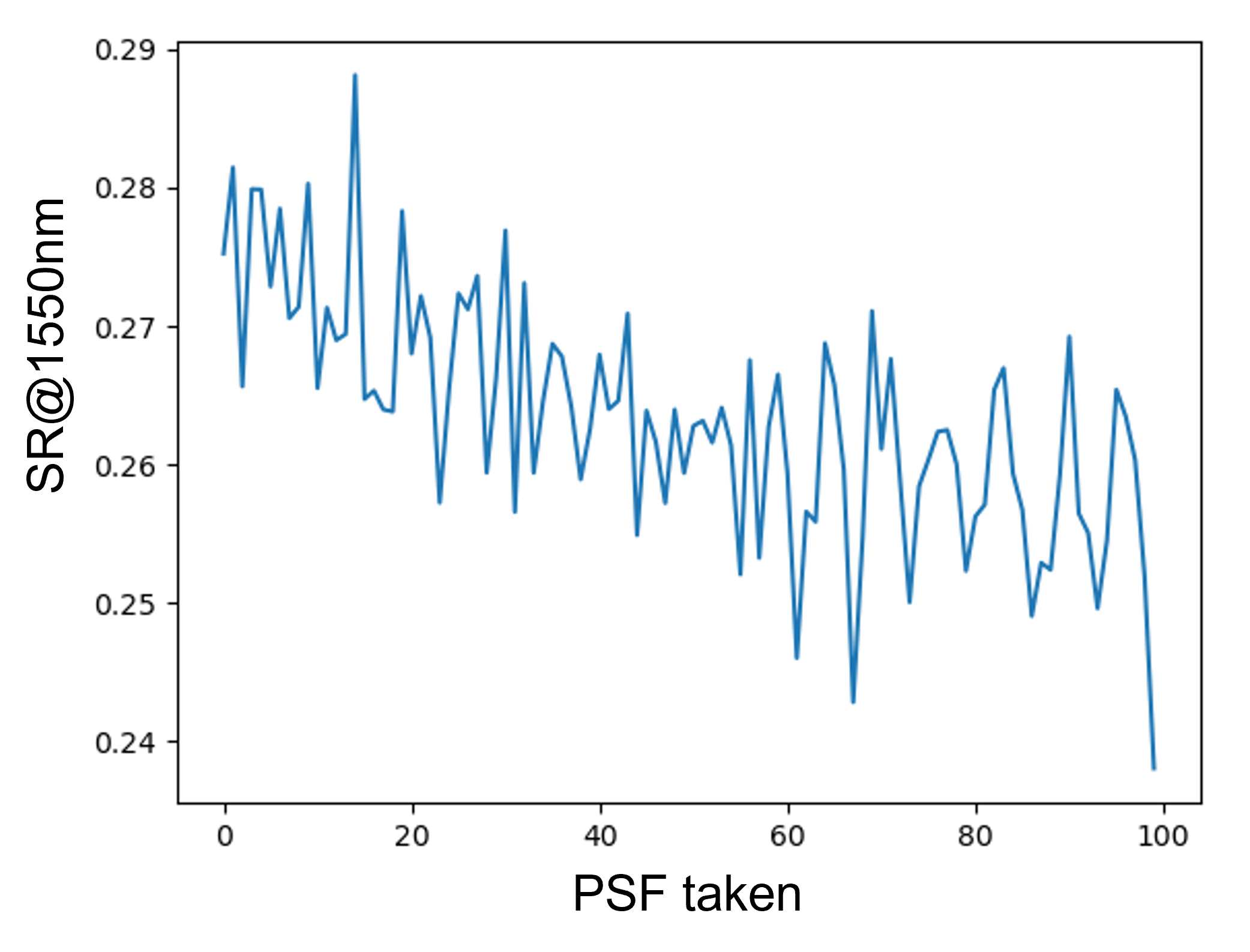}
    \caption{Repeatability test of SR measurement. Each SR measurement took 27s to acquire. Full repeatability test was performed over 50 minutes. }
    \label{fig:repeatability_test}
\end{figure}

To estimate the Strehl ratio, we measured the PSF during a full turn of the phaseplate (27 seconds), so as to ensure some repeatability of the measurement. A repeatability test is performed when it appears that there is some drift over time in the performances of the bench (see Figure \ref{fig:repeatability_test}). 

 We concluded from this test that the systematic error on SR measurement is $\sigma_{SR}=0.83\%$, because we have not been able to determine when this drift appears.
 
 \subsection{Bench test results}

\begin{figure}
    \centering
    \includegraphics[width=0.45\linewidth]{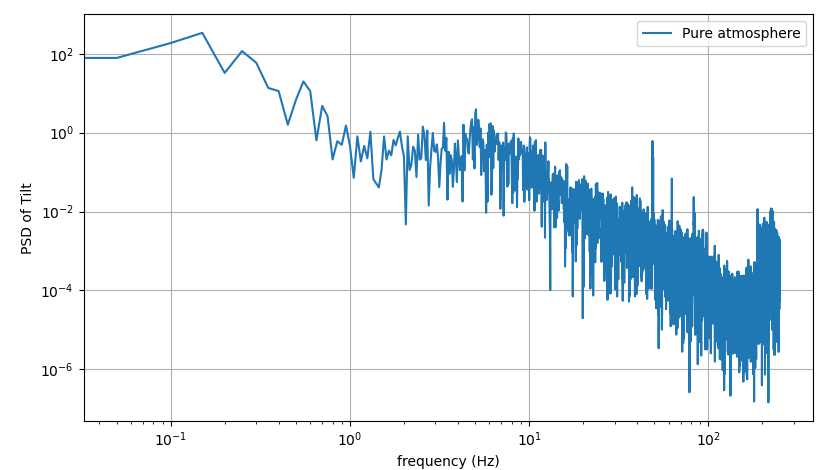}
    \includegraphics[width=0.45\linewidth]{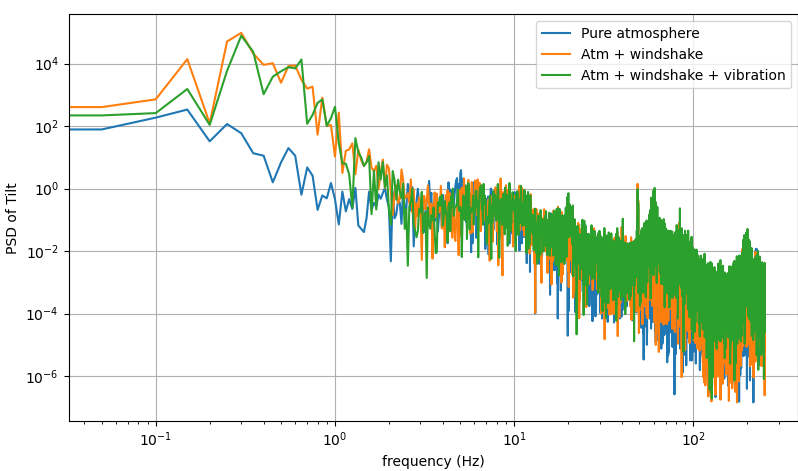}
    \caption{PSD of tilt measured on bench. \textbf{Left :} PSD created by the phase plate (= atmosphere). We see a vibration at 48 Hz. This PSD has been measured on 8 different dasy and seems permanent. \textbf{Right : } PSD of tilt of atmosphre, atmosphere + windshake, atmosphere + windshake + vibration}
    \label{fig:PSD_bench}
\end{figure}

\begin{figure}
    \centering
    \includegraphics[width=0.9\linewidth]{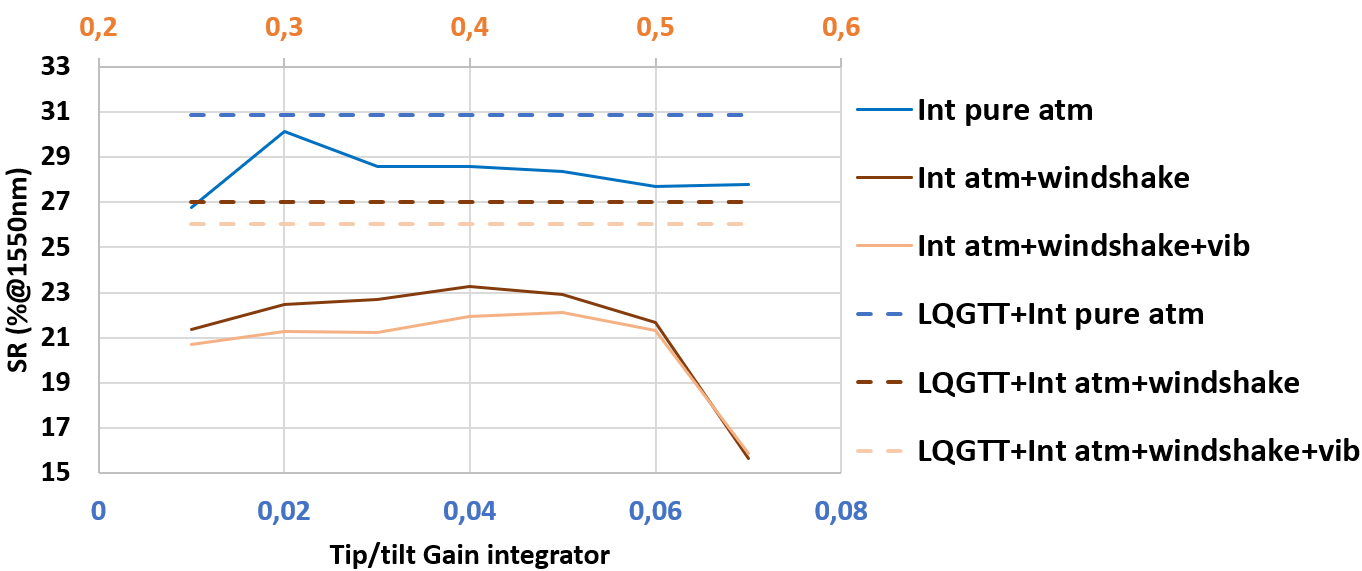}
    \caption{On-bench performance. We see that LQG has better results than the integrator, especially in presence of windshake and vibrations}
    \label{fig:result_bench_perf}
\end{figure}

We measured the tip-tilt POL on the bench. We plot the tilt the pure atmosphere then we added windshake and tip tilt. We see on Figure \ref{fig:PSD_bench} that in the supposedly pure atmosphere case that there is already a vibration on the bench. This explains the optimal integrator gain seen on Figure \ref{fig:result_bench_perf} in the supposedly pure atmosphere case, where the optimal gain is extremely low. 

We plot on Figure \ref{fig:result_bench_perf} the result to compare the different controller. We plot the integrator in function of the gain used to find the best performances. LQG having two parameters to optimise we plot the best result we could get with fudge factor and model order optimisation. The optimal parameters found are summed up in Table \ref{tab:LQG_parameters}. We see an evolution of the parameter order similar to simulation, but as seen in Figure \ref{fig:PSD_bench}, the model order need to be pretty high from the start due to vibrations. A very high fudge factor is common when measurement noise does not dominate error sources, which is the case in the current bench configuration.   

\begin{table}[ht]
    \centering
    \begin{tabular}{|c|c|c|}
        \hline
        Atmospheric condition & fudgefactor & Model order  \\
         \hline
        pure atmosphere & 100 & 12\\
          \hline
        atmosphere + windshake & 100 & 23\\
          \hline
         atmosphere + windshake + vibrations & 1000 & 42 \\ 
          \hline 
    \end{tabular}
    \caption{Optimal parameters for LQG test}
    \label{tab:LQG_parameters}
\end{table}

We see that the LQG restores the performances close to the situation with only the pure atmosphere in presence of both windshake and vibration. When looking at the integrator gain that gives the best performances, the gain for the pure atmosphere case is very low (0.02), while the gain is in the expected range for atm+windshake and atm+windshake+vib (0.4 and 0.45 respectively). We see in table \ref{tab:LQG_parameters} the LQG parameters. We see a larger model order than for simulation (14, 23 and 42 compared to 5,15 and 25 in simulation). 
By analysing the PSD of the phase measured in closed loop we see a vibration around 48H which explains these oddities. Fudgefactor is high due to the low noise present on the bench. The system is dominated not by detector noise, but by other sources n the bench like non-linearity, which we can adapt the LQG to by increasing fudgefactor. 

\begin{figure}
    \centering
    \includegraphics[width=0.9\linewidth]{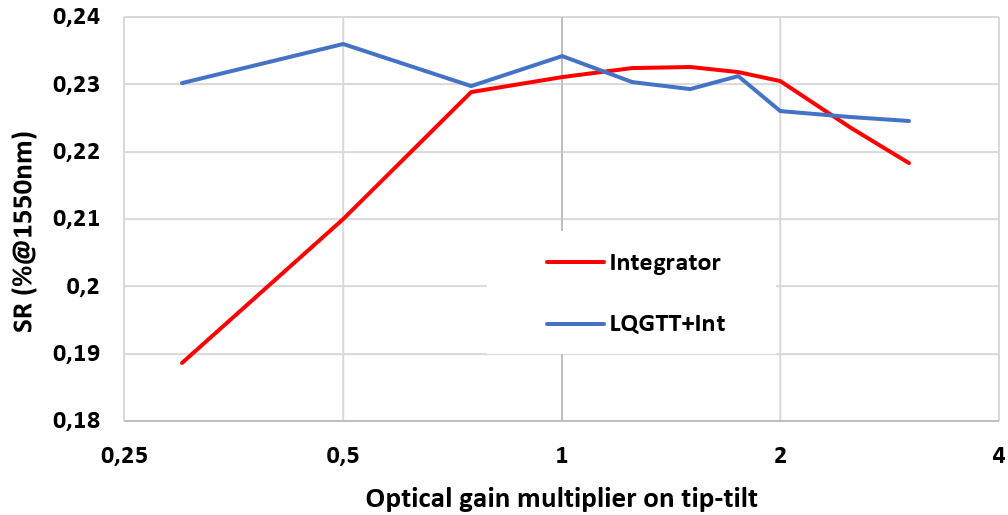}
    \caption{SR variation on bench when Tip-tilt optical gains are incorrect (here due to multiplication)}
    \label{fig:OG_effect_bench}
\end{figure}

A second test was carried out to check the effect of bad OG estimation on the controllers on bench. We multiply the OG of the Tip-Tilt modes only. The NCPA compensation was not used for this phase to avoid the NCPA catastrophe described in Chambouleyron et al 2020 \cite{chambouleyron_pyramid_2020} (so optimal SR are lower for this test). Result are shown in Figure \ref{fig:OG_effect_bench} 

We see result similar to the simulated results. While the LQG sees small effect of a change of optical gain between 0.3 and 3, the integrator performances are greatly impacted. 


\section{Conclusion}
Windshake and vibrations are a big problem for the ELT as it greatly reduces its performances. To compensate this new perturbations we implement a more advanced controller and adapt it to the PyWFS and test it on the MICADO SCAO bench. LQG control allows the MICADO SCAO to reach its performance goals despite the ELT specific disturbances in simulation as well as on bench. We also show its resilience to optical gain variations, a very encouraging property. 

Automation steps are under way for the LG to adapt automatically to the changes of condition in the atmosphere. This includes recomputing the atmosphere model at regular intervals, the frequency of this recomputation being one of the parameters to optimise. Furthermore, optimisation of parameters such as the fudge factor and the perturbation model order also needs to be automated. This was already tested for GTC \cite{marquis_first_2024}.
Another development to come is expanding the LQG controller beyond Tip-Tilt, since it is very likely that vibrations will appear on additional low-order modes, such as astigmatism.

\acknowledgments 
 
This work has been supported by PEPR Origins-UPCAO project, funded by the French Research National Agency (France 2030 investment plan), grant No. ANR-22-EXOR-0017. This work has benefited from the support of the French Programme d’Investissement d’Avenir through the project F-CELT ANR-21-ESRE-0008, the CNRS 80 PRIME program, the CNRS INSU IR budget, the Action Spécifique Haute Résolution Angulaire (ASHRA) of CNRS/INSU co-funded by CNES, the Observatoire de Paris, the Ile de France region (DIM ACAV/ACAV+ and ORIGINES), the LIRA and LCF laboratories.

\bibliography{biblio_proceeding} 
\bibliographystyle{spiebib} 

\end{document}